\documentclass{jfm}

\usepackage{makecell}
\usepackage{graphicx,subfig}
\usepackage{newtxtext,newtxmath}
\usepackage{natbib}

\usepackage{algorithm}
\usepackage{algpseudocode}
\usepackage{caption}

\usepackage{hyperref}
\usepackage[nameinlink,capitalize]{cleveref}
\hypersetup{
  colorlinks=true,
  linkcolor=blue,
  citecolor=blue,
  urlcolor=blue,
}

\crefname{figure}{figure}{figures}
\Crefname{figure}{Figure}{Figures}
\crefname{table}{table}{tables}
\Crefname{table}{Table}{Tables}
\crefname{equation}{equation}{equations}
\Crefname{equation}{Equation}{Equations}

\title{Symbolic identification of tensor equations in multidimensional physical fields}

\author{Tianyi Chen\aff{1}, 
Hao Yang\aff{1},
Wenjun Ma\aff{1},\and
Jun Zhang\aff{1}\corresp{\email{jun.zhang@buaa.edu.cn}}}

\affiliation{\aff{1}School of Aeronautic Science and Engineering, Beihang University, Beijing, 100191, China}

\begin{document}
\maketitle

\begin{abstract}
Recently, data-driven methods have shown great promise for discovering governing equations from simulation or experimental data. However, most existing approaches are limited to scalar equations, with few capable of identifying tensor relationships. In this work, we propose a general data-driven framework for identifying tensor equations, referred to as Symbolic Identification of Tensor Equations (SITE). The core idea of SITE—representing tensor equations using a host–plasmid structure—is inspired by the multidimensional gene expression programming (M-GEP) approach. To improve the robustness of the evolutionary process, SITE adopts a genetic information retention strategy. Moreover, SITE introduces two key innovations beyond conventional evolutionary algorithms. First, it incorporates a dimensional homogeneity check to restrict the search space and eliminate physically invalid expressions. Second, it replaces traditional linear scaling with a tensor linear regression technique, greatly enhancing the efficiency of numerical coefficient optimization. We validate SITE using two benchmark scenarios, where it accurately recovers target equations from synthetic data, showing robustness to noise and small sample sizes. Furthermore, SITE is applied to identify constitutive relations directly from molecular simulation data, which are generated without reliance on macroscopic constitutive models. It adapts to both compressible and incompressible flow conditions and successfully identifies the corresponding macroscopic forms, highlighting its potential for data-driven discovery of tensor equation.
\end{abstract}

\begin{keywords}
machine learning
\end{keywords}


\section{Introduction}
\label{sec:intro}

Physical phenomena in the real world are governed by fundamental laws. Traditionally, these laws have been derived from first principles through deductive reasoning and formulated as mathematical expressions, i.e., governing equations. However, as systems become increasingly complex, this traditional first-principles approach faces significant challenges. Recent breakthroughs in artificial intelligence (AI) are revolutionizing the natural sciences, driving a paradigm shift from first-principles modeling to data-driven approaches. In particular, symbolic regression has gained increasing attention for its ability to identify analytical expressions of governing equations directly from data\citep{bruntonMachineLearningFluid2020,wangScientificDiscoveryAge2023}.

Long before the explosive development of AI, the concept of symbolic regression had already been embodied in evolutionary algorithms. For instance, genetic programming (GP) constructs symbolic expression trees to fit data by simulating biological evolution \citep{kozaGeneticProgrammingMeans1994}. \citet{bongardAutomatedReverseEngineering2007}, as well as \citet{schmidtDistillingFreeFormNatural2009}, pioneered the application of GP concepts to data-driven equation discovery, successfully constructing governing equations for low-dimensional systems from experimental data. Subsequent advances in GP-based methodologies have yielded notable equation discovery frameworks, including EPDE \citep{maslyaevDatadrivenPartialDerivative2019}, SGA-PDE \citep{chenSymbolicGeneticAlgorithm2022}, and PySR \citep{cranmerInterpretableMachineLearning2023}. Another evolutionary algorithm used for symbolic regression is gene expression programming (GEP) \citep{ferreiraGeneExpressionProgramming2001a}. Unlike GP, GEP employs fixed-length linear strings to control expression complexity and utilizes open reading frames to improve expression diversity. GEP have been applied to the construction of Reynolds stress models \citep{weatherittNovelEvolutionaryAlgorithm2016,zhao2020rans}, subgrid-scale models for large eddy simulations \citep{liDatadrivenModelDevelopment2021}, and non-equilibrium constitutive equations \citep{maDimensionalHomogeneityConstrained2024}. In parallel, a distinct class of symbolic regression approaches based on sparse regression has emerged. A landmark contribution is the sparse identification of nonlinear dynamics (SINDy) method developed by \citet{bruntonDiscoveringGoverningEquations2016}. SINDy transforms the discovery of ordinary differential equations (ODEs) into a sparse regression problem by identifying dominant terms from a library of nonlinear functions. Subsequent developments in this direction include the use of constrained Galerkin projection \citep{loiseauConstrainedSparseGalerkin2018}, the application of sparse regression to macroscopic hydrodynamic systems \citep{zhangDatadrivenDiscoveryGoverning2020}, the incorporation of weak-form formulations to expand candidate libraries \citep{Gurevich_Golden_Reinbold_Grigoriev_2024}, and the inference of hidden stochastic dynamics from high-dimensional data \citep{luWeakCollocationRegression2024}.

While data-driven equation discovery has made significant strides, studies have pointed out that purely data-driven methods are highly sensitive to data quality and quantity \citep{chenPhysicsinformedLearningGoverning2021,reinboldRobustLearningNoisy2021,cohenPhysicsinformedGeneticProgramming2024}. Such methods often suffer from overfitting and yield expressions that lack physical interpretability. To address these issues, researchers have incorporated physical constraints and prior knowledge into symbolic regression. For instance, \citet{udrescuAIFeynmanPhysicsinspired2020} guided the search using physical inductive biases; \citet{bakarjiDimensionallyConsistentLearning2022a}, \citet{fukamiDatadrivenNonlinearTurbulent2024}, and \citet{maDimensionalHomogeneityConstrained2024} incorporated dimensional analysis, greatly enhancing the physical interpretability of the results; \citet{chenPhysicsinformedLearningGoverning2021} combined physics-informed neural networks (PINNs) with sparse regression.

Despite these advances, most existing methods focus on discovering scalar equations. As system dimension increases, the candidate function space in scalar regression methods grows exponentially. For instance, consider a time-invariant system in Cartesian coordinates. If only positions, velocities, and their first-order spatial derivatives are considered, the number of independent variables jumps from 3 in one dimension to 15 in three dimensions. This exponential growth of the search space greatly diminishes the efficiency of equation identification.

A natural solution is to extend symbolic regression from discovering scalar equations to identifying tensor equations. Tensors inherently capture directional dependencies and exhibit transformation invariance (e.g., under translations, rotations, and reflections), enabling more compact and physically meaningful representations of high-dimensional systems. In many physical domains, such as fluid mechanics, most physical quantities are intrinsically tensors. Therefore, the symbolic identification of tensor equations offers not only greater expressive power but also a more physically appropriate modeling framework.

At present, general-purpose methods for symbolic tensor equation discovery are still lacking. One representative approach is the multidimensional GEP (M-GEP) algorithm, which extends evolutionary strategies to tensor systems \citep{weatherittNovelEvolutionaryAlgorithm2016}. Another approach is the Cartesian tensor-based sparse regression (CTSR), which applies sparse regression techniques to identify tensor equations \citep{zhangCTSRCartesianTensorbased2025}. In addition, other methods such as RUDEs \citep{lennonScientificMachineLearning2023} and SPIDER \citep{Gurevich_Golden_Reinbold_Grigoriev_2024,goldenPhysicallyInformedDatadriven2023}, incorporate translational invariance and symmetry into their frameworks.

In this work, we propose a general data-driven framework for identifying tensor equations, termed Symbolic Identification of Tensor Equations (SITE). SITE offers a broadly applicable approach for identifying tensor governing equations across diverse domains. Inspired by M-GEP, SITE encodes tensor expressions using a host–plasmid architecture but adopts a different evolutionary strategy to enhance robustness. To address the challenges of large, redundant search spaces and the generation of physically meaningless expressions, SITE incorporates dimensional homogeneity check during evolutionary process. Additionally, a tensor linear regression (TLR) technique is introduced to replace the conventional linear scaling step \citep{keijzerImprovingSymbolicRegression2003} used in scalar symbolic regression, significantly accelerating convergence and improving numerical accuracy.

The remainder of this paper is organized as follows. §\ref{sec:method} introduces the SITE methodology in detail. §\ref{sec:benchmark} presents numerical experiments on two benchmark problems to validate the robustness and accuracy of SITE. §\ref{sec:discovery} explores the discovery of constitutive relations from molecular simulation data, and §\ref{sec:conclusion} summarizes our key conclusions.

\section{Methodology}
\label{sec:method}
\begin{figure} 
    \centering  
    \includegraphics[width=13.5cm]{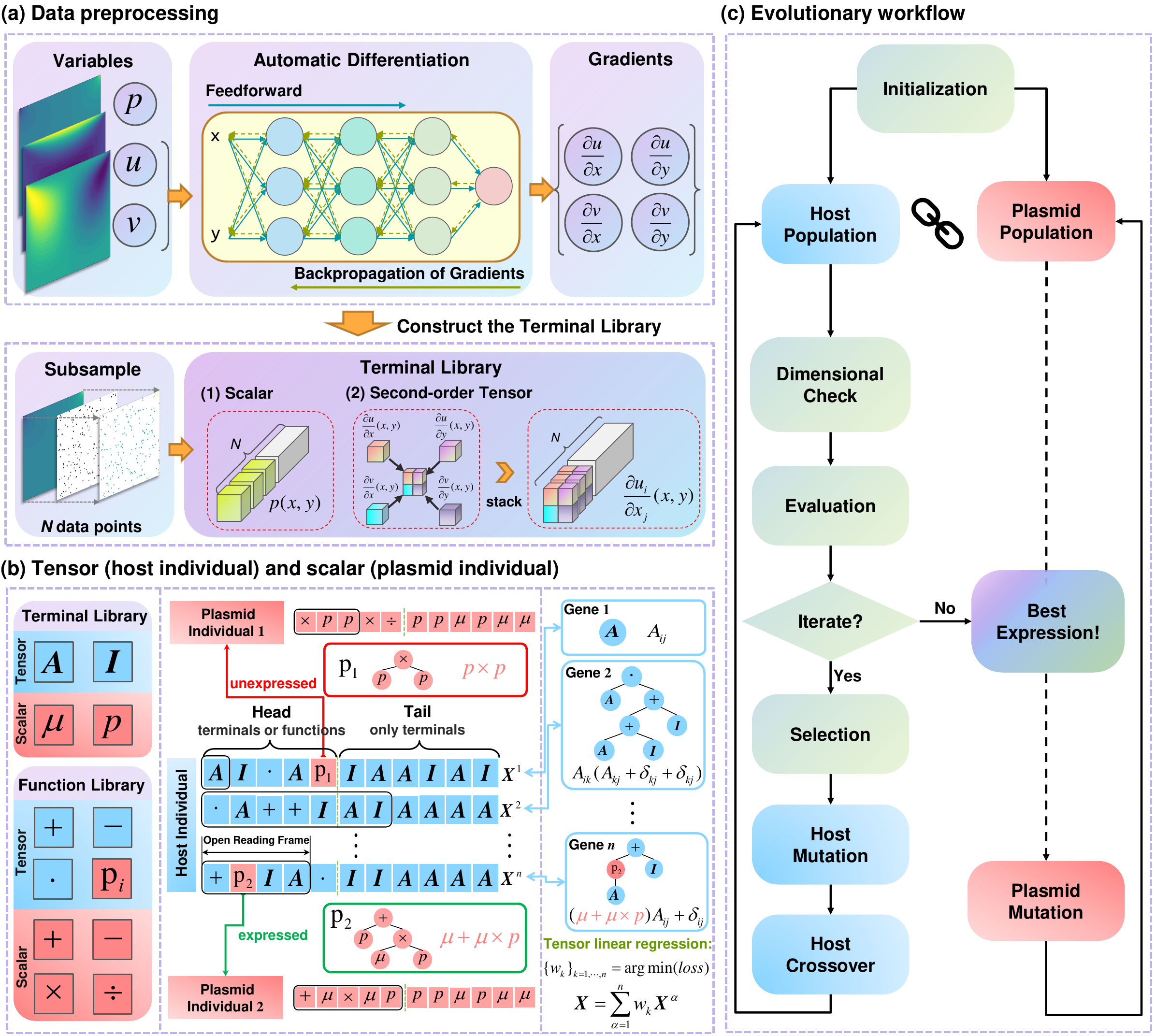}
    \caption{Overview of the SITE framework. (a) Data preprocessing, including the calculation of the gradients of macroscopic quantities and construction of the terminal library. (b) Schematic diagram of tensor (host individual) and scalar (plasmid individual). (c) Flowchart of the evolutionary workflow.} 
    \label{fig1}
\end{figure}

This section presents the methodological design of SITE. As illustrated in \cref{fig1}, the process of symbolic identification using SITE begins with data preparation and preprocessing. A terminal library for both tensors and scalars is then constructed based on these physical quantities. Using this library and a predefined set of symbolic operators, SITE generates candidate equations by composing various symbolic structures, which are iteratively optimized through an evolutionary workflow.

We elaborate on the three core components of SITE that enable the identification of tensor equations with both physical interpretability and quantitative accuracy. The evolutionary process serves as the foundation for identifying tensor equation structures. In parallel, dimensional homogeneity check and tensor linear regression ensure the physical interpretability and quantitative accuracy of the identified results, respectively.

\subsection{Evolutionary process of SITE}
\subsubsection{Conventional GEP}
SITE implements the evolutionary process based on the conventional GEP and extends it to handle tensor equations through a nested host–plasmid architecture. To provide the necessary context, we begin with a brief overview of the conventional GEP algorithm.

In conventional GEP, an initial population of individuals—each representing a mathematical expression—is constructed. Each individual, or chromosome, is composed of one or more linear sequences called genes. These genes are formed by combining elements from a terminal library (containing variables and constants) and a function library (containing mathematical operators). Each gene consists of a head of fixed length $h$,  which may include functions and terminals, and a tail of length $t = h \times ({a_{\max }}) + 1$, restricted to terminals only. Here, ${a_{\max }}$ means the max arity (the number of arguments a mathematical function takes) of the functions in the candidate function library. Each gene can be parsed into an expression tree (ET) using a deterministic decoding scheme, and the final expression represented by a chromosome is obtained by linking the outputs of multiple ETs using a predefined linking function (e.g., addition). Having multiple genes within a single chromosome enhances the representational power and exploratory capabilities.

After the initial population is generated, the fitness of each individual in the population is evaluated. The mean relative error is a common loss function used for evaluation, which is defined as

\begin{equation}
    {\cal L}(Y,\hat Y) = \frac{1}{N}\sum\limits_{k = 1}^N {\left| {\frac{{{{\hat Y}_k} - {Y_k}}}{{{Y_k}}}} \right|} ,
\end{equation}
where ${\hat Y_k}$ is the $k$-th predicted value of all the $N$ data points and ${Y_k}$ is the ground truth.

Selection then proceeds using a hybrid strategy combining elitism and tournament selection, as designed in \citet{ferreiraGeneExpressionProgramming2001a}. 
 Elites are retained directly, while others undergo genetic operations like mutation, transposition, and crossover. For a detailed description of the genetic operators, refer to \cref{appA}. The evolutionary cycle repeats until either the loss of an expression falls below a predefined threshold or the maximum number of evolution generations is reached.

\subsubsection{Extension to tensor equations based on M-GEP}
We now extend the conventional GEP to handle tensor equations, focusing on second-order Cartesian tensors. The nested host–plasmid architecture used in SITE is adapted from M-GEP. SITE enhances the evolutionary strategy and generalizes the framework to support a broader applications in tensor equation identification.

Assume the general form of the tensor equation to be identified is
\begin{equation}
    T_{ij} = F(\{\boldsymbol{a}_{ij}\},\{\boldsymbol{x}\})
,
\end{equation}
where $T_{ij}$ is the target second-order tensor, while $\boldsymbol{a}_{ij}$ and $\boldsymbol{x}$ represent a group of second-order tensors and scalars, respectively. Our goal is to identify the underlying symbolic form $F(\cdot)$ from data.

To handle mixed-order tensor operations and incorporate scalars into tensor expressions, we adopt a host–plasmid architecture in SITE. Specifically, host individuals are used to represent second-order tensor expressions, while plasmid individuals, which exist in dependency on their host individuals, are used to embed scalars into the tensor expressions.

As shown in \cref{fig1}(b), the host gene uses a special ‘p’ operator to invoke a corresponding plasmid, enabling scalar-tensor coupling. In both the terminal library and the function library, we need to prepare two parts, corresponding respectively to the hosts (the blue part in the diagram) and the plasmids (the red part in the diagram). The host terminal library includes candidate second-order tensors, while its function library contains tensor operators with equal input and output tensor orders, such as tensor addition, tensor multiplication, and inner product, as well as the ‘p’ operator used to invoke embedded scalar expressions. In contrast, the plasmid terminal library contains candidate scalars, and its function library includes basic scalar operators.

When a ‘p’ operator appears in a host gene, it triggers a scalar multiplication between the tensor below the ‘p’ in the ET and the scalar expression generated by the corresponding plasmid. For example, the last gene of the host individual in \cref{fig1}(b) has a head of length 5 and a tail of length 6. Here, the open reading frame (ORF) determines which elements of a gene are used to construct the ET. Starting from the head of the gene, the ORF sequentially reads elements (functions or terminals) from left to right until it naturally terminates due to the gene’s structural constraints (such as the end of the tail or grammatical rules), ensuring that the generated ET is complete. The ORF and the tensor expression of this gene ($X_{ij}^{n}$) are:
\begin{equation}
    \begin{array}{l}
\left[ {\begin{array}{*{20}{c}}
 + &{{{\text{p}}_2}}&{{\delta _{ij}}}&{{A_{ij}}}
\end{array}} \right],\quad
X_{ij}^n = {{\text{p}}_2}({A_{ij}}) + {\delta _{ij}},
\end{array}
\end{equation}
respectively, where ‘$\text{p}_2$’ will evoke the plasmid individual 2. The scalar expression represented by plasmid individual 2 is
\begin{equation}
    \mu  + (\mu  \times p),
\end{equation}
and the corresponding complete expression should be
\begin{equation}
    X_{ij}^n = (\mu  + \mu  \times p){A_{ij}} + {\delta _{ij}}.
\end{equation}

\begin{figure} 
\centering  
\includegraphics[width=13.5cm]{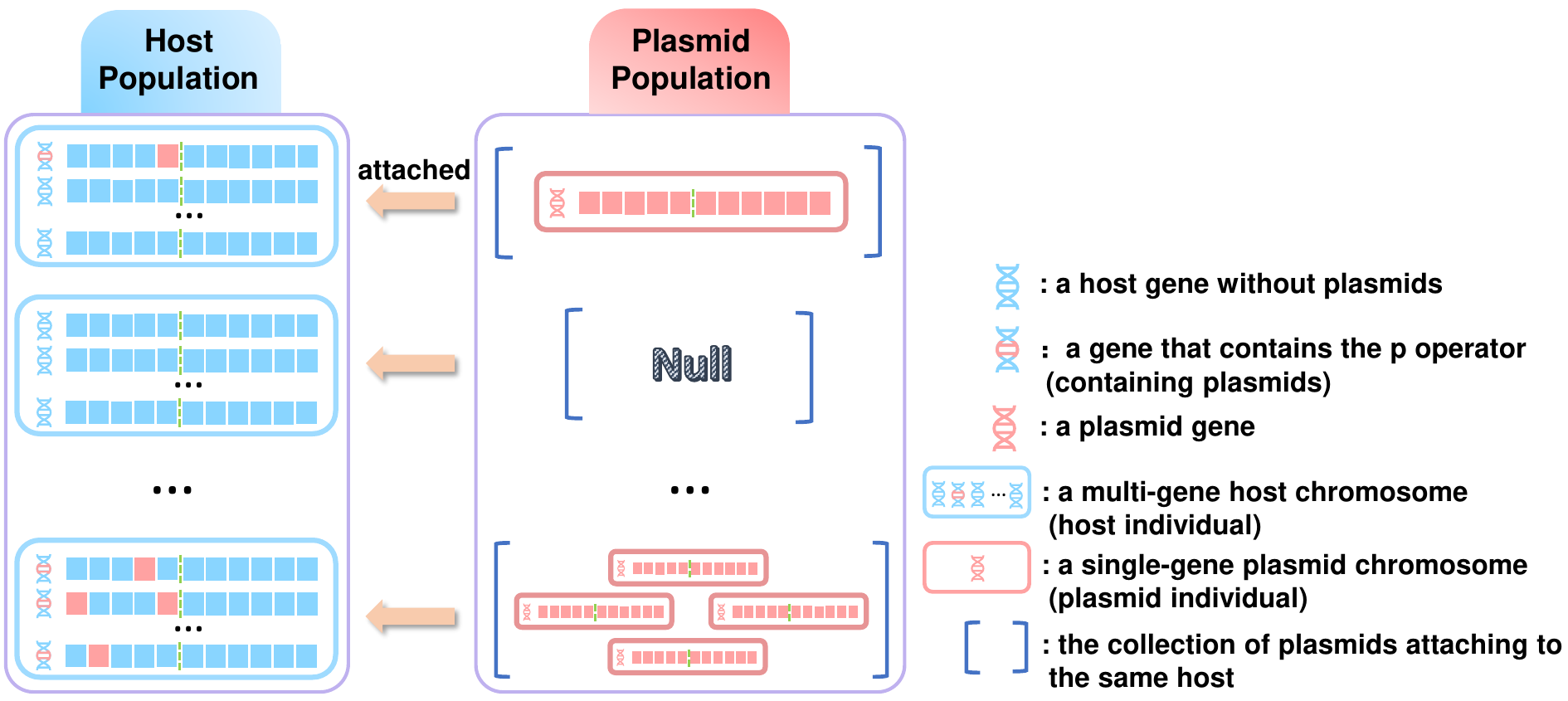}
\caption{Schematic diagram of generating plasmid populations based on host population.} 
\label{fig2}
\end{figure}

In the specific implementation of SITE, the plasmid population construction is dynamic. The number of ‘p’ operators in a host chromosome determines how many plasmid individuals are attached to it. As illustrated in \cref{fig2}, for each ‘p’ operator, a plasmid individual is randomly generated at initialization, and all the plasmid individuals constitute the plasmid population. Essentially, the core evolutionary process is carried out by the host population with a fixed number of individuals, while the plasmid population is merely an appendage to the host population. The number of individuals in the plasmid population fluctuates in accordance with changes in the host population. Note that SITE retains unexpressed plasmids temporarily suppressed by genetic operations rather than destroying them. This strategy enhances robustness by retaining genetic diversity and preventing the premature loss of potentially useful sub-expressions.

The evolutionary process of SITE (see \cref{fig1}(c)) is similar to that of conventional GEP. In each generation, the host population is evaluated and selected based on fitness, guiding the progression of the evolutionary process. In contrast, the plasmid population does not undergo direct evaluation or selection. Instead, these plasmid individuals serve to assist in the evaluation of the host population, indirectly contributing to the evolutionary process. After selection, the host offspring will execute both modification and crossover operations, while the plasmid population evolves solely through modification. This design retains the linkage integrity between plasmids and their hosts, thereby avoiding disruptive recombination.

Since the target variable changes from a scalar to a tensor, we also need to design a loss function suitable for tensor symbol regression. In SITE, the loss function used for evaluation is defined as
\begin{equation}
    {\cal L}({y_{ij}},{\hat y_{ij}}) = \sum\limits_{i = 1}^{{n_{\dim }}} {\sum\limits_{j = 1}^{{n_{\dim }}} {\frac{{\left\| {{{\hat y}_{ij}} - {y_{ij}}} \right\|_2^2}}{{\left\| {{y_{ij}}} \right\|_2^2}}} } ,
    \label{lossfunc}
\end{equation}
where ${y_{ij}} \in {\mathbb{R}^{{n_{{\text{data}}}}}}$ and ${\hat y_{ij}} \in {\mathbb{R}^{{n_{{\text{data}}}}}}$ denote the value sequences across all data points for the $(i,j)$-th component of the target tensor $\boldsymbol{Y}$ and the predicted tensor $\boldsymbol{\hat Y}$, respectively. $\left\| {{{\hat y}_{ij}} - {y_{ij}}} \right\|_2^2$ represents the squared \( L^2 \) norm of the residual corresponding to the tensor component in a specific direction, while $\left\| {{y_{ij}}} \right\|_2^2$ represents the squared \( L^2 \) norm of the target tensor component. This loss function measures the average relative error across all tensor components. After several generations of evolution, the optimal tensor expression is identified.

\subsection{Dimensional homogeneity check and seed injection strategy}
In SITE, we introduce a fundamental physical constraint—dimensional homogeneity—into the equation discovery by performing dimensional homogeneity check. We select the seven base physical quantities of the SI-unit System (i.e., mass ($M$), length ($L$), time ($T$), electric current ($I$), thermodynamic temperature ($\Theta $), amount of substance ($N$), and luminous intensity ($J$)) as the basic dimension. All known physical quantities’ dimensions can be derived by combining the above base dimensions.

\begin{figure} 
\centering  
\includegraphics[width=13.5cm]{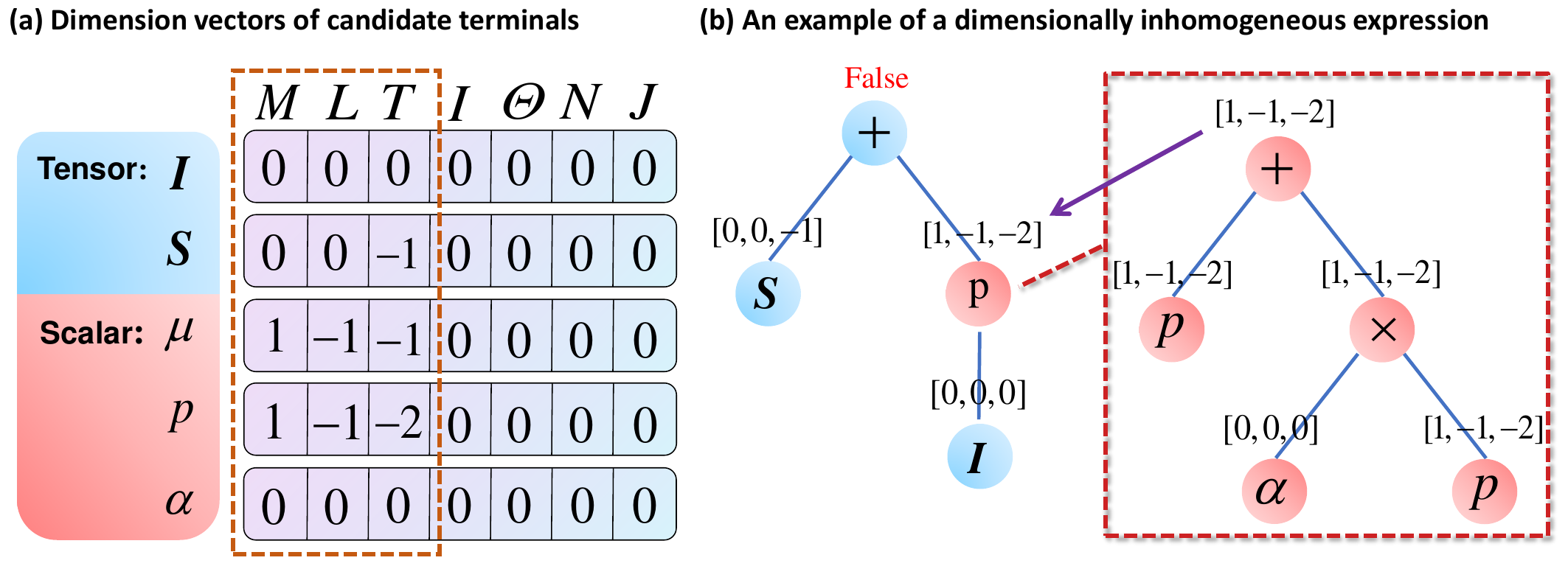}
\caption{Schematic diagram of dimension vectors and dimensional homogeneity check.} 
\label{fig3}
\end{figure}

\Cref{fig3}(a) presents the dimension vectors (DVs) used for dimensional analysis, where the specific numbers represent the exponents of the corresponding dimensions. In SITE, every element in the candidate terminal set is assigned a unique DV, which serves as an intrinsic attribute of the terminal. We determine the DV at each node, as well as the resulting DV of the entire expression, by traversing the ET from the bottom up. If the expression itself involves operations that violate the dimensional homogeneity constraint as shown in \cref{fig3}(b), a summation occurs between two nodes with different DVs, the individual is immediately classified as dimensionally inhomogeneous. If the expression is dimensional homogeneous, its resulting DV is then compared to that of the target variable; if the two DVs are identical, the individual is considered to satisfy the dimensional homogeneity constraint.

In the SITE framework, the dimensional homogeneity check is performed not only for scalar quantities but also extended to two additional cases involving tensors. First, for operations between tensors, since all components within a tensor share the same dimension, addition and inner products follow the same rules as in the scalar case. Second, for operations between tensors and scalars, SITE only involves scalar multiplication between a tensor and a scalar, following the scalar multiplication rules. Details of dimensional homogeneity check for scalar expressions using DVs refer to \citet{reissmannConstrainingGeneticSymbolic2025}.

In the evaluation step of the evolutionary workflow, for individuals that fail the dimensional homogeneity check, the loss function is not applied to compute their loss value. Instead, they are assigned a sufficiently large loss, thus gradually eliminating expressions without physical meaning in the evolutionary process. The physical constraint method used in SITE is a hard constraint, which completely ensures that the resulting equations satisfy the dimensional homogeneity. 

Note that increasing the number and length of host and plasmid genes to represent complex tensor expressions can sharply reduce the proportion of dimensionally consistent individuals, making it difficult for the population to generate valid expressions even after many generations. To mitigate this, we propose a seed injection strategy, inspired by population diversity enhancement mechanisms in evolutionary computation \citep{goldbergGeneticAlgorithmsSharing1987}. Specifically, at fixed intervals during evolution, several host genes that individually satisfy the dimensional homogeneity constraint are manually generated and concatenated to form a single seed individual. Multiple copies of this individual are then injected into the host population. This strategy facilitates the emergence of dimensionally homogeneous expressions in early generations. Additionally, it helps the population escape local optima, analogous to how invasive species reshape ecological systems.

\subsection{Tensor linear regression technique}
Note that conventional GEP excels at exploring the functional form of equations but struggles with determining accurate numerical coefficients. This limitation arises because coefficient optimization in conventional GEP relies solely on the mutation of random numerical constants (RNCs), a highly stochastic process that results in low efficiency in optimizing coefficients. To address this issue, the linear scaling technique \citep{keijzerImprovingSymbolicRegression2003} has been introduced into conventional GEP. This approach adjusts the output of candidate expressions through a global scaling factor, allowing for a better fit to the target data. This method proves effective in scalar contexts, as each individual adjusts all numerical values with a single scaling factor.

In tensor symbolic regression, using a single global scaling factor cannot effectively fit all components of a predicted tensor expression. To address this, we introduce a tensor linear regression (TLR) technique, which assigns an independent coefficient to each gene and fits the predicted tensor expression to the target through a linear combination of tensor features. This approach enhances coefficient optimization efficiency and enables implicit feature selection by assigning different importance to different genes.

The TLR task in SITE is formulated as follows: For a given target tensor ${Y_{ij}}$ and a set of tensor features and a set of tensor features $\left( {\begin{array}{*{20}{c}}
{{{\left( {{X_{ij}}} \right)}_1}}& \cdots &{{{\left( {{X_{ij}}} \right)}_{{n_{{\text{gene}}}}}}}
\end{array}} \right)$ expressed by host genes, where $i,j \in \{ 1,2, \cdots ,{n_{{\text{dim}}}}\} $ are tensor indices corresponding to the physical spatial dimensions and $n_{\text{gene}}$ denotes the number of host genes, we need to prepare their values at all ${n_{{\text{data}}}}$ points. Given the values of all tensors at each data point, we can define the target tensor as
\begin{equation}
    {y_{ij}} = \left( {\begin{array}{*{20}{c}}
{{{\left( {{Y_{ij}}} \right)}_1}}\\
 \vdots \\
{{{\left( {{Y_{ij}}} \right)}_{{n_{{\text{data}}}}}}}
\end{array}} \right)
\end{equation}
and the set of tensor features as
\begin{equation}
    {\Theta _{ij}} = \left( {\begin{array}{*{20}{c}}
    {{{\left( {{X_{ij}}} \right)}_{11}}}& \ldots &{{{\left( {{X_{ij}}} \right)}_{1{n_{{\text{gene}}}}}}}\\
     \vdots & \ddots & \vdots \\
    {{{\left( {{X_{ij}}} \right)}_{{n_{{\text{data}}}}1}}}& \cdots &{{{\left( {{X_{ij}}} \right)}_{{n_{{\text{data}}}}{n_{{\text{gene}}}}}}}
    \end{array}} \right).
\end{equation}
Our objective is to seek a coefficient vector $\boldsymbol{w}$ such that the predicted $\hat{y}_{ij}$ fits the target $y_{ij}$ and minimizes the loss function. The loss function using in evaluation has been defined as (\ref{lossfunc}). Here, the complete loss function in TLR is
\begin{equation}
    {\cal L}({y_{ij}},{\hat y_{ij}}) = \sum\limits_{i = 1}^{{n_{\dim }}} {\sum\limits_{j = 1}^{{n_{\dim }}} {\frac{{{{({{\hat y}_{ij}} - {y_{ij}})}^T}({{\hat y}_{ij}} - {y_{ij}})}}{{{y_{ij}}^T{y_{ij}}}}} } .
    \label{summation}
\end{equation}
The summation in (\ref{summation}) means that every component of a tensor is taken into account. The predicted $\hat{y}_{ij}$ is obtained by a weighted linear combination of $n_{\text{gene}}$ features:
\begin{equation}
    \hat{y}_{ij}
    = \left( {\begin{array}{*{20}{c}}
    {{{\left( {{{\hat Y}_{ij}}} \right)}_1}}\\
     \vdots \\
    {{{\left( {{{\hat Y}_{ij}}} \right)}_{{n_{{\text{data}}}}}}}
    \end{array}} \right) = {\Theta _{ij}}w = \left( {\begin{array}{*{20}{c}}
    {{{\left( {{X_{ij}}} \right)}_{11}}}& \ldots &{{{\left( {{X_{ij}}} \right)}_{1{n_{{\text{gene}}}}}}}\\
     \vdots & \ddots & \vdots \\
    {{{\left( {{X_{ij}}} \right)}_{{n_{{\text{data}}}}1}}}& \cdots &{{{\left( {{X_{ij}}} \right)}_{{n_{{\text{data}}}}{n_{{\text{gene}}}}}}}
    \end{array}} \right)\left( {\begin{array}{*{20}{c}}
    {{w_1}}\\
     \vdots \\
    {{w_{{n_{{\text{gene}}}}}}}
    \end{array}} \right).
    \label{2.10}
\end{equation}

The coefficient vector will be determined from
\begin{equation}
    \boldsymbol{w} = \arg \min {\cal L}({y_{ij}},{\hat y_{ij}}).
\end{equation}
By taking the partial derivative of the loss function and setting it to zero, an analytical solution for the coefficient vector $\boldsymbol{w}$ can be obtained as
\begin{equation}
    w = {\left( {\sum\limits_{i = 1}^{{n_{\dim }}} {\sum\limits_{j = 1}^{{n_{\dim }}} {\frac{{\Theta _{ij}^T{\Theta _{ij}}}}{{y_{ij}^T{y_{ij}}}}} } } \right)^{ - 1}}\left( {\sum\limits_{i = 1}^{{n_{\dim }}} {\sum\limits_{j = 1}^{{n_{\dim }}} {\frac{{\Theta _{ij}^T{y_{ij}}}}{{y_{ij}^T{y_{ij}}}}} } } \right).
    \label{2.12}
\end{equation}
For the analytical form of the coefficient vector to be valid, there is a prerequisite condition that the coefficient matrix must be nonsingular. This condition is essentially equivalent to the features in the feature set ${\Theta _{ij}}$ being linearly independent. Therefore, to apply (\ref{2.12}) for solving the coefficient vector, we need to first perform feature selection on the feature set, selecting all the linearly independent features to construct a reduced feature set ${\Theta _{ij}}^\prime $ and ensuring that it is column-linearly independent.

With the coefficient vector $\boldsymbol{w}$, we change the predicted expression from the summation of each feature tensor to (\ref{2.10}) before evaluation. The entire TLR process takes place after dimensional homogeneity check and before evaluating the final loss, serving as the final optimization before an individual is evaluated. The coefficient vector of each individual is independent of its sub-ETs and is computed only before evaluation without participating in any subsequent genetic operations.

\section{Validation on benchmark problems}
\label{sec:benchmark}
We validate and demonstrate the capabilities of SITE in identifying tensor equations using two benchmark scenarios. These two cases correspond to well-established models in mathematical physics, namely the Maxwell stress tensor and the Reynolds-stress transport equation. In addition, we also evaluate the significant performance improvement of SITE after incorporating the TLR technique, as well as the framework’s robustness to data noise and dataset size. 
\subsection{Maxwell stress tensor}
We first consider the Maxwell stress tensor in the context of electromagnetism. The Maxwell stress tensor, typically defined as
\begin{equation}
    {T_{ij}} = {\varepsilon _0}\left( {{E_i}{E_j} - \frac{1}{2}{E_k}{E_k}{\delta _{ij}}} \right) + \frac{1}{{{\mu _0}}}\left( {{B_i}{B_j} - \frac{1}{2}{B_k}{B_k}{\delta _{ij}}} \right),
    \label{3.1}
\end{equation}
describes the distribution of electromagnetic momentum and forces within a medium. Here, $E_i$ and $B_i$ are the components of the electric and magnetic fields, respectively. $\varepsilon _0$ is the vacuum permittivity and $\mu_0$ is the vacuum permeability.

\begin{figure} 
\centering  
\includegraphics[width=13.5cm]{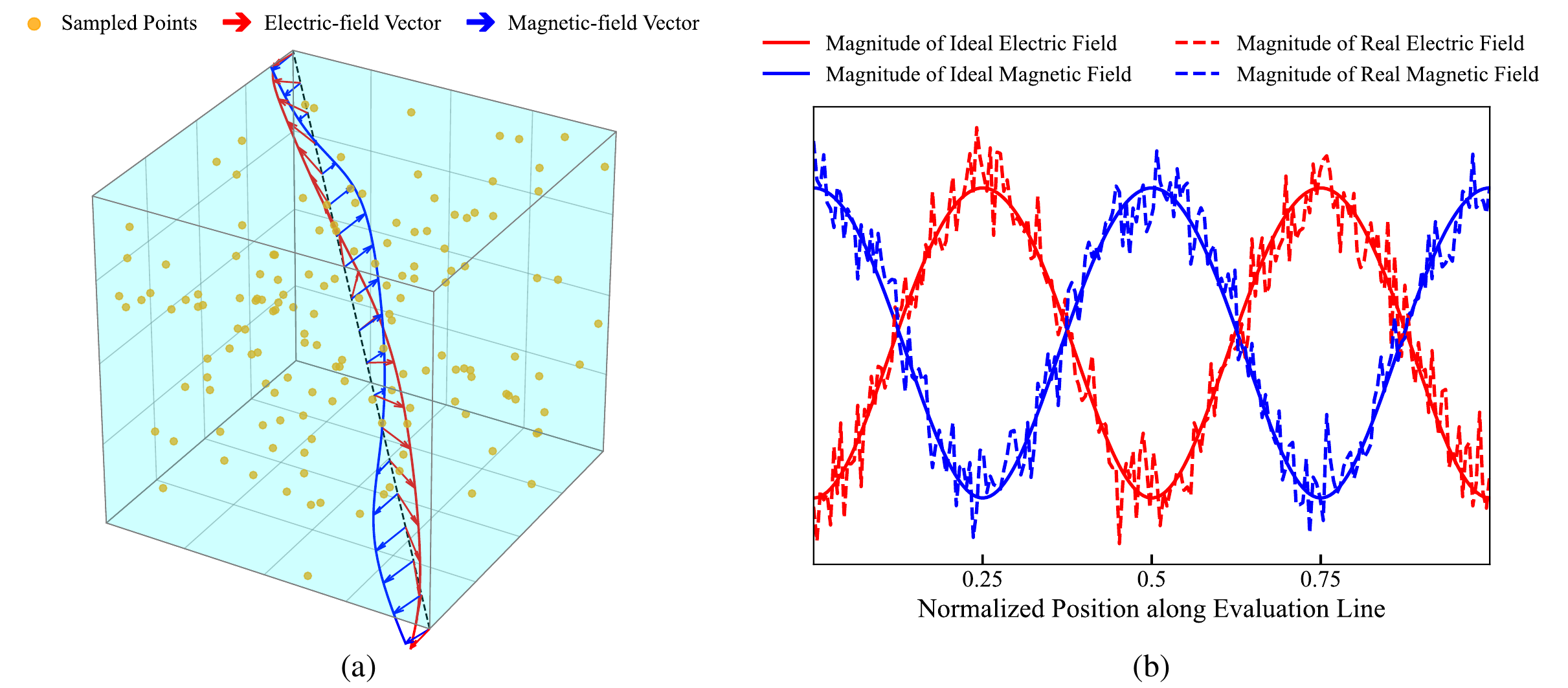}
\caption{(a) Schematic diagram of the electric field, the magnetic field and sampled points. For clarity of illustration, only the vectors along the guiding dashed line are shown. (b) Trend lines of the electromagnetic field magnitude at the points along the guiding line in (a) before and after adding $5\%$ Gaussian noise. The solid line represents the ideal setting, and the dashed line represents the real situation.} 
\label{fig4}
\end{figure}

A three-dimensional cube with sinusoidally varying fields is used to evaluate the behavior governed by (\ref{3.1}). The electric field is given by
\begin{equation}
    E(x,y,z) = \left[ {{E_0}\sin (\pi x),{E_0}\cos (\pi y),{E_0}\sin (\pi z)} \right],
\end{equation}
and the magnetic field is given by
\begin{equation}
    B(x,y,z) = \left[ {{B_0}\cos (\pi x),{B_0}\sin (\pi y),{B_0}\cos (\pi z)} \right],
\end{equation}
where ${E_0} = {10^6}{\text{ V/m}}$ and ${B_0} = {10^{ - 3}}{\text{ T}}$ are fixed amplitudes. To construct the dataset, 150 spatial points are randomly sampled within the cube ($- 1\le x,y,z\le1$), and the corresponding Maxwell stress tensor   is computed analytically using (\ref{3.1}). The schematic diagram of this case is shown in \cref{fig4}(a).

In this case, the target variable is defined as $T_{ij}$. For tensors, the candidate function library and the candidate terminal library are defined as $\{  + , - , \cdot ,{\text{p}}\} $ and $\{ {E_i}{E_j},{B_i}{B_j},{\delta _{ij}}\} $, respectively, while for scalars, the candidate function library and the candidate terminal library are defined as $\{  + , - , \times , \div \} $ and $\{ {E_{kk}},{B_{kk}},{\varepsilon _0},{B_0}\} $. The candidate terminal libraries of different tests are exactly the same.

To highlight the contribution of TLR to improving coefficient optimization efficiency, we conduct comparative experiments under three configurations: (1) using only RNCs, (2) using only TLR, and (3) combining TLR with RNCs. When TLR is not employed, RNCs must be introduced into the scalar candidate terminal library to form numerical constants in expressions. These RNCs are randomly sampled floating-point numbers within the range $[-10,10]$. The hyperparameter settings for the comparative experiments are basically the same, except that additional genetic operations need to be performed on RNCs when they are introduced (see \Cref{appA}). Detailed parameter settings for the SITE algorithm, such as the maximum number of evolutionary generations and the number of host individuals, are provided in \cref{appB}. The derived equations for this test, along with the number of iterations and the CPU runtime (measured on a system equipped with a 13th Gen Intel® Core™ i9-13900K CPU at 3.00 GHz, 32 GB RAM, and running a 64-bit operating system), are summarized in \cref{tab:maxwell}.

\begin{table}
  \begin{center}
\def~{\hphantom{0}}
  \begin{tabular}{cccc}
      TLR used  & RNCs used   &   Derived equation & \makecell{Iterations \\ (time)} \\[8pt]
       No   & Yes & ${T_{ij}} = 0.970{\varepsilon _0}{E_i}{E_j} - 0.133{\varepsilon _0}{E_k}{E_k}{\delta _{ij}}$ & \makecell{ $2000$ \\ ($1666\text{s}$)}\\[6pt]
       Yes   & No & ${T_{ij}} = {\varepsilon _0}{E_i}{E_j} - 0.5{\varepsilon _0}{E_k}{E_k}{\delta _{ij}} + \frac{1}{{{\mu _0}}}{B_i}{B_j} - 0.5\frac{1}{{{\mu _0}}}{B_k}{B_k}{\delta _{ij}}$ &\makecell{ $714$ \\ ($580\text{s}$)}\\[6pt]
       Yes  & Yes & ${T_{ij}} = {\varepsilon _0}{E_i}{E_j} - 0.5{\varepsilon _0}{E_k}{E_k}{\delta _{ij}} + \frac{1}{{{\mu _0}}}{B_i}{B_j} - 0.5\frac{1}{{{\mu _0}}}{B_k}{B_k}{\delta _{ij}}$ &\makecell{ $23$ \\ ($20\text{s}$)}\\[6pt]
  \end{tabular}
  \caption{Performance comparison of SITE using different coefficient optimization configurations.}
  \label{tab:maxwell}
  \end{center}
\end{table}

It can be seen from \cref{tab:maxwell} that regardless of whether TLR is used, SITE is capable of discovering complex tensor equations from the data. However, we also observe that without TLR, the framework fails to identify a completely correct equation even after reaching the maximum number of evolutionary generations (2000 iterations). Specifically, the coefficients of discovered terms are incorrect and the combination of ${B_i}{B_j}$, ${B_k}{B_k}{\delta _{ij}}$, and $\frac{1}{{{\mu _0}}}$ is not identified. This is because optimization based on RNCs relies only on completely random mutations to alter the numerical coefficients in the expression. In contrast, when TLR is used, the correct equation is accurately identified within just 23 iterations. It is evident that TLR substantially improves the efficiency and accuracy of tensor equation identification. Moreover, we tested a third configuration where RNCs are removed and only TLR is used. SITE still successfully discovers the exact equation within 714 iterations, demonstrating that TLR alone is sufficient to achieve both high accuracy without relying on RNCs. Since the mingling of RNCs with symbolic terminals sometimes may hinder the generation of concise expressions, all cases in this article use only TLR to optimize coefficients unless otherwise specified.

To evaluate the robustness of SITE to noise, we test its performance under three noise levels: $5\%$, $10\%$, and $20\%$. Specifically, as illustrated as \cref{fig4}(b), Gaussian noise is first added to the electric and magnetic field strengths to generate perturbed physical quantities. Then, Maxwell stress tensors are computed from these noisy fields and further perturbed by Gaussian noise of the same level. These noisy data are used for the subsequent tests. Here, we use both RNCs and TLR here to get results quickly.

\begin{table}
  \begin{center}
\def~{\hphantom{0}}
  \begin{tabular}{cccc}
      Noise level  &   Derived equation & Error \\[8pt]
       $5\%$  & ${T_{ij}} = 0.999{\varepsilon _0}{E_i}{E_j} - 0.500{\varepsilon _0}{E_k}{E_k}{\delta _{ij}} + 1.007\frac{1}{{{\mu _0}}}{B_i}{B_j} - 0.499\frac{1}{{{\mu _0}}}{B_k}{B_k}{\delta _{ij}}$ & $0.25 \pm 0.27\% $\\[6pt]
       $10\%$  & ${T_{ij}} = 0.998{\varepsilon _0}{E_i}{E_j} - 0.500{\varepsilon _0}{E_k}{E_k}{\delta _{ij}} + 1.014\frac{1}{{{\mu _0}}}{B_i}{B_j} - 0.501\frac{1}{{{\mu _0}}}{B_k}{B_k}{\delta _{ij}}$ & $0.45 \pm 0.55\% $\\[6pt]
       $20\%$  & ${T_{ij}} = 0.997{\varepsilon _0}{E_i}{E_j} - 0.500{\varepsilon _0}{E_k}{E_k}{\delta _{ij}} + 1.032\frac{1}{{{\mu _0}}}{B_i}{B_j} - 0.513\frac{1}{{{\mu _0}}}{B_k}{B_k}{\delta _{ij}}$ & $1.53 \pm 1.40\% $ \\[6pt]
  \end{tabular}
  \caption{Equations identified by SITE under different levels of Gaussian noise ($5\%$, $10\%$, and $20\%$).}
  \label{tab:noise}
  \end{center}
\end{table}

The derived equations with different levels of noise are listed in \cref{tab:noise}. The error in the table is defined as the average relative error of the identified non-zero coefficients with respect to the ground truth. When Gaussian noise is added, the coefficients in the equation identified by SITE deviate increasingly from the ground truth as the noise level increases, and the error grows almost linearly with the noise level. Even with up to $20\%$ Gaussian noise, SITE still successfully identifies the complete structure of the expression, although there are some deviations in the coefficients. This proves that the framework has great robustness against artificial noise.

\subsection{Reynolds-stress transport equation}
\label{sec:RTE}
In many physical systems, the evolution of tensors is governed by time-dependent differential equations. To evaluate the capability of SITE to identify such dynamic tensor equations, we consider the following general form of a time-dependent tensor evolution equation:
\begin{equation}
    \frac{{d{T_{ij}}}}{{dt}} = F(\{\boldsymbol{a}_{ij}\},\{\boldsymbol{x}\})
\end{equation}
where $T_{ij}$ is a second-order tensor, while $\boldsymbol{a}_{ij}$ and $\boldsymbol{x}$ represent a group of second-order tensors and scalars, respectively. Note that when spatial derivatives of the tensor $T_{ij}$ appear on the right-hand side of the equation, it becomes a partial differential equation.

To generate data for the discovery of this kind of tensor equation, simulations are performed for decaying homogeneous isotropic turbulence. Argon gas is used as the working fluid, with initial conditions of $273.15{\text{ K}}$ temperature, $100,000{\text{ Pa}}$ pressure, and zero mean velocity. The initial turbulent kinetic energy is set to $4265.9{\text{ }}{{\text{m}}^2}{\text{/}}{{\text{s}}^2}$, corresponding to an initial turbulent Mach number of 0.3. All simulations are carried out using the open-source finite volume solver OpenFOAM \citep{jasakOpenFOAMLibraryComplex2007}. A Crank–Nicolson scheme is used for time integration with a fixed time step of $1 \times {10^{ - 8}}{\text{ s}}$, and the total simulation time is set to $1 \times {10^{ - 6}}{\text{ s}}$. Under these conditions, the Reynolds stress transport equation degenerates into
\begin{equation}
    \frac{{d{R_{ij}}}}{{dt}} =  - \frac{2}{3}\varepsilon {\delta _{ij}},
    \label{3.5}
\end{equation}
where $\varepsilon$ is the isotropic dissipation rate.

In this case, the dataset consists of the Reynolds stress tensor $R_{ij}$, its temporal derivatives $\frac{{d{R_{ij}}}}{{dt}}$, the dissipation rate $\varepsilon$, and the turbulent kinetic energy $k$, recorded at 100 time instances uniformly sampled from $1 \times {10^{ - 8}}{\text{ s}}$ to $1 \times {10^{ - 6}}{\text{ s}}$. Since the Reynolds stress tensor in this case depends only on time and exhibits simple temporal variation, its gradient can be accurately computed using straightforward finite differences. To investigate how the number of data points used influences the SITE results, we conduct comparative experiments using all data points, 75 data points, 50 data points, and 25 data points. When utilizing only a subset of the data, we randomly sample the number of data points from the full dataset for training. This random process is determined by a specified random seed. We use 500 different random seeds for sampling, and the final coefficients are obtained from the statistics of these 500 training runs. The target variable is defined as $\frac{{d{R_{ij}}}}{{dt}}$. For tensors, the candidate function library and the candidate terminal library are defined as $\{  + , - , \cdot ,{\text{p}}\} $ and $\{ {R_{ij}},{\delta _{ij}}\} $, respectively, while for scalars, the candidate function library and the candidate terminal library are defined as $\{  + , - , \times , \div \} $ and $\{ k,\varepsilon \} $. Note that $R_{ij}$ and $k$ in the terminal library do not appear in (\ref{3.5}), and they are introduced as distraction terms.

\begin{table}
  \begin{center}
\def~{\hphantom{0}}
  \begin{tabular}{cc}
       \makecell{Dataset size \\(The percentage of the total dataset)}  &  Derived equation \\[8pt]
       $100 (100\%)$  & $\frac{{d{R_{ij}}}}{{dt}} =  - 0.6617\varepsilon {\delta _{ij}}$\\[6pt]
       $75 (75\%)$  & $\frac{{d{R_{ij}}}}{{dt}} =  - (0.6617 \pm 0.0001)\varepsilon {\delta _{ij}}$ \\[6pt]
       $50 (50\%)$  & $\frac{{d{R_{ij}}}}{{dt}} =  - (0.6617 \pm 0.0002)\varepsilon {\delta _{ij}}$ \\[6pt]
       $25 (25\%)$  & $\frac{{d{R_{ij}}}}{{dt}} =  - (0.6617 \pm 0.0004)\varepsilon {\delta _{ij}}$ \\[6pt]       
  \end{tabular}
  \caption{Equations identified by SITE from datasets of different sizes.}
  \label{tab:datasize}
  \end{center}
\end{table}

The derived equations are listed in \cref{tab:datasize}. It can be seen that the derived equation matches the target equation very well, even there are several distraction variables in the candidate terminal library. This case also demonstrates the capability of SITE in identifying time-dependent governing equations. Moreover, we observe that the overall average of the obtained equations remains highly stable as the dataset size varies. Examining the standard deviation of each equation coefficient reveals that smaller datasets result in greater fluctuations in the coefficients. When the dataset is too small, the model may overfit anomalous data points. Therefore, we propose that appropriate multiple subsampling and averaging the results from each sampling can effectively enhance the stability of the results.

\section{Equation identification based on molecular simulation}
\label{sec:discovery}
We now apply SITE to high-fidelity data generated via direct simulation Monte Carlo (DSMC). Unlike the data of benchmark cases in §\ref{sec:benchmark}, the DSMC datasets are generated directly from microscopic particle interactions, without embedding any macroscopic governing equations such as the Navier–Stokes equations or predefined constitutive relations. Our goal is to demonstrate that SITE can discover the macroscopic tensor constitutive relation from molecular simulation data generated without imposing any macroscopic constitutive assumptions. Additionally, we evaluate SITE’s ability to identify the corresponding macroscopic forms under two distinct flow conditions and assess its robustness in capturing small-magnitude terms.

\subsection{Data generation and preprocessing}
\label{sec:data}
In this section, we employ the DSMC method \citep{birdMolecularGasDynamics1994} to generate flow field data, following the approach in \citet{maDimensionalHomogeneityConstrained2024} and \citet{zhangDatadrivenDiscoveryGoverning2020}. Since the aim of this section is to discover tensor macroscopic governing equations of fluid dynamics, it is crucial that the data be independent of macroscopic modeling assumptions. DSMC fulfills this requirement by simulating particle dynamics without reliance on predefined continuum-level equations such as the Navier–Stokes (NS) equations. In DSMC, each simulation molecule statistically represents a large number of real molecules. From these molecular-level dynamics, macroscopic physical quantities are obtained via statistical averaging. For instance, the velocity field ($v_i$) and the stress field ($\sigma_{ij}$) are computed as
\begin{equation}
    {v_i} = \frac{1}{{{N_{\text{p}}}}}\sum\limits_{{\text{cell}}} {{c_i}} ,\quad{\sigma _{ij}} = \frac{{mF}}{{{V_{{\text{cell}}}}}}\left( {\sum\limits_{{\text{cell}}} {{c_i}} {c_j} - {N_{\text{p}}}{v_i}{v_j}} \right),
\end{equation}
where $N_p$ is the number of simulation molecules in the sampling cell, $c_i$ is the velocity of the $i$-th particle, $m$ is the molecular mass, and $F$ is the simulation ratio in DSMC. With the data generated from DSMC, we can construct a mapping from spatial coordinates to various macroscopic variables. As illustrated in \cref{fig1}(a), at each given data point $(x,y)$, we associate corresponding physical quantities such as density $\rho(x,y)$, horizontal velocity $u(x,y)$ and vertical velocity $v(x,y)$.

In our study, two primary sources of noise affect the accuracy of equation discovery. One is the numerical noise arising from the computation of spatial gradients, and the other is the statistical noise inherent in particle-based methods such as DSMC. To mitigate the first type of noise, we apply the neural-network-based automatic differentiation technique to calculate gradients of physical variables. Specifically, an artificial neural network is trained to map spatial coordinates to macroscopic variables, and gradients of arbitrary order are obtained by backpropagating through the network. This allows accurate evaluation of spatial derivatives at any point in the domain. As for the statistical noise introduced by particle simulation, we can quantitatively estimate how the statistical error changes with the number of statistical samples \citep{hadjiconstantinouStatisticalErrorParticle2003} with
\begin{equation}
    {E_Q} = \frac{1}{{\sqrt {{N_0}M} }}\frac{{\sqrt {\left\langle {{Q^2}} \right\rangle } }}{{\left| {\left\langle Q \right\rangle } \right|}},
\end{equation}
where $E_Q$ represents the statistical error of quantity $Q$, $\left\langle {{\cdot}} \right\rangle$ means the statistical mean value, $N_0$ is the number of particles in a statistical cell, and $M$ is the number of independent samples. In our study, we set the ${N_0}M$ of our simulation to $4 \times {10^9}$, which means the statistical error of $\sigma_{ij}$ is around ${10^{ - 5}}$. An error of this magnitude is acceptable for our experiments.

Once all physical variables and their gradients are obtained, we subsample a portion of the data points from the computational domain for discovering the governing equation. The rationale for subsampling has been discussed in §\ref{sec:RTE}. As shown in the lower part of \cref{fig1}(a), the sampled physical quantities are organized into structured arrays, which serve as the terminal library corresponding to each problem.

\subsection{Constitutive equation discovery under different flow conditions}
\label{sec:4.2}
We simulate lid-driven cavity flows using the DSMC method described in §\ref{sec:data}. The simulated domain is a square cavity with a side length of $L = 1{\text{ m}}$, filled with argon gas. All four walls maintain at a constant temperature of $273.15 \text{ K}$. The bottom, left, and right walls are stationary, while the top wall (lid) moves horizontally at a constant speed. The number of grids is set to $400 \times 400$. Each grid cell provides DSMC statistical outputs of the horizontal and vertical velocities, the four components of stress, the temperature, and the static pressure.

Two flow conditions are considered by setting the lid velocity to different values. In the first case, the lid speed is set to  , corresponding to a low-speed, nearly incompressible flow. In the second case, the lid velocity is increased to  , resulting in a transonic, compressible flow. The DSMC simulations employ the variable hard sphere (VHS) model, in which the dynamic viscosity varies with temperature according to
\begin{equation}
    \mu  = {\mu _{{\text{ref}}}}{\left( {\frac{T}{{{T_{{\text{ref}}}}}}} \right)^\omega }
\end{equation}
where ${\mu _{{\text{ref}}}} = 2.117 \times {10^{ - 5}}{\text{ kg/(m}} \cdot {\text{s)}}$ denotes the reference viscosity coefficient, ${T_{{\text{ref}}}} = 273.15{\text{ K}}$ denotes the reference temperature, and $\omega  = 0.81$ denotes the viscosity exponent. The global Knudsen number in both cases is fixed at ${\text{Kn}} = 0.005$.

In constructing the candidate function library, we introduced a set of basic tensor and scalar operators. As for the candidate terminal library, special consideration was given to the physical relevance of stress modeling. Since constitutive relation of stress reflects the transport of momentum in the flow field—and such transport is primarily driven by velocity gradients—we included physical quantities related to the velocity gradient tensor. The velocity gradient tensor ${D_{ij}} = \partial {u_i}/\partial {x_j}$ can be decomposed into a symmetric and an antisymmetric part, given as
\begin{equation}
    {D_{ij}} = {S_{ij}} + {\Omega _{ij}},
\end{equation}
where ${S_{ij}} = \frac{1}{2}\left( {{D_{ij}} + {D_{ji}}} \right)$ is the symmetric strain-rate tensor, and ${\Omega _{ij}} = \frac{1}{2}\left( {{D_{ij}} - {D_{ji}}} \right)$ is the antisymmetric rotation-rate tensor. In both cases, the target variable is defined as $\sigma_{ij}$. For tensors, the candidate function library is defined as $\{  + , - , \cdot ,{\text{p}}\} $ and $\{ {S_{ij}},{\delta _{ij}}\} $, respectively, while for scalars, the candidate function library and the candidate terminal library are defined as $\{  + , - , \times , \div \} $ and $\{ {D_{kk}},p,\mu \} $. For a detailed description of the construction of the candidate term library used in this study, please refer to \cref{appC}.

To enable the computation of velocity gradients, we trained neural networks to learn the mapping from spatial coordinates to the two velocity components within the flow field. Specifically, we used all $400\times400$ data points from the computational domain as the training set. The mapping was modeled using a fully connected feedforward neural network comprising 8 hidden layers, each with 60 neurons and tanh activation functions. The training process was carried out in two stages. First, we used the Adam optimizer \citep{kingmaAdamMethodStochastic2017} for $20,000$ iterations to achieve a reasonable initial solution. Then, the optimization was refined using the L-BFGS-B \citep{byrdLimitedMemoryAlgorithm1995} optimization until convergence. This hybrid optimization strategy ensures both stability and precision in capturing smooth and differentiable velocity fields suitable for gradient extraction via automatic differentiation.

\begin{figure} 
\centering  
\includegraphics[width=12cm]{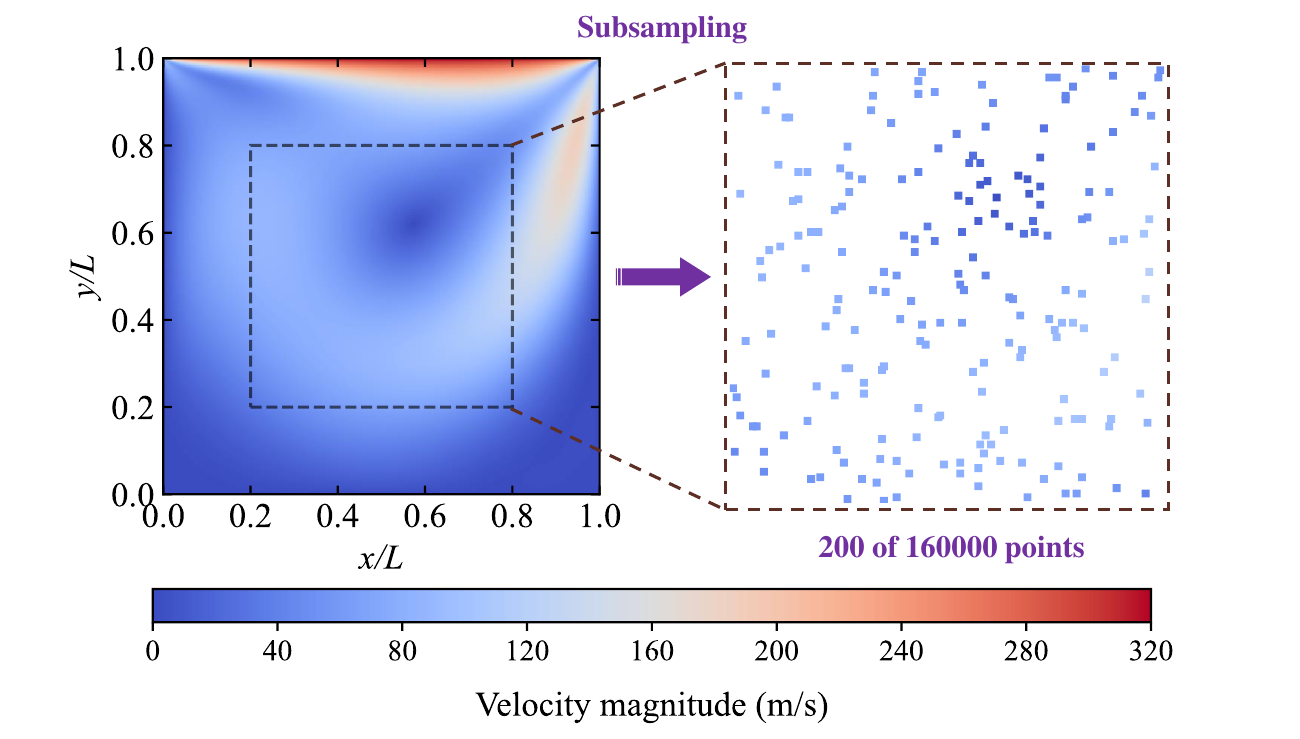}
\caption{Distribution of velocity magnitude in the compressible cavity flow case and the spatial locations of sampled data points used for SITE.} 
\label{fig5}
\end{figure}

The last step before applying SITE is subsampling the dataset. To mitigate the influence of numerical noise when computing gradients near the boundaries, we restricted the sampling region to the central portion of the domain—specifically, the $20\%$ to $80\%$ range along both spatial directions. From this subregion, 200 data points were randomly selected and used for symbolic regression and equation discovery. \Cref{fig5} illustrates the distribution of velocity magnitude (defined as $\sqrt {{u^2} + {v^2}} $) across the domain in the compressible cavity flow case, along with a schematic representation of the sampled data points used for training. This selection strategy ensures that the discovered equations are based on high-quality gradient information.

For reference, the complete Newtonian constitutive relation is given by
\begin{equation}
    {\sigma _{ij}} =  - 2\mu {S_{ij}} + \frac{2}{3}\mu {D_{kk}}{\delta _{ij}} + p{\delta _{ij}}.
\end{equation}
This form can be derived under the continuum hypothesis. The first term is the shear stress term, which arises from the internal friction caused by the relative sliding (shear deformation) between fluid layers, corresponding to the anisotropic behavior of shear stress. The second term is the compressible term, accounting for the viscous resistance generated during volumetric changes. This term is typically negligible in incompressible flows where the velocity divergence is close to zero. The third term is the static pressure term, representing the isotropic compressive stress imposed by the thermodynamic pressure, which is independent of the flow state.

\begin{table}
  \begin{center}
\def~{\hphantom{0}}
  \begin{tabular}{lc}
      Lid velocity (m/s)  & Derived equation \\[8pt]
       50   & ${\sigma _{ij}} =  - 1.906\mu {S_{ij}} + 1.000p{\delta _{ij}}$\\[6pt]
       337   & ${\sigma _{ij}} =  - 1.914\mu {S_{ij}} + 0.676\mu {D_{kk}}{\delta _{ij}} + 1.000p{\delta _{ij}}$\\
  \end{tabular}
  \caption{Identified equations of cavity flow.}
  \label{tab:cavity}
  \end{center}
\end{table}

As shown in \cref{tab:cavity}, we summarize the constitutive equations discovered in the two test cases. To ensure robustness, we subsample the data using different random seeds. For each case, several independent experiments were conducted with varying data points. The terms presented in the table are those that appear frequently across these experiments, and the corresponding coefficients are computed as averages over the respective runs. In the compressible case, the compressible viscous term is consistently identified, whereas this term is rarely detected in the low-speed incompressible case. This indicates that SITE not only successfully discovers constitutive relations from molecular simulation data but also adapts the discovered form according to the underlying flow characteristics.

In certain experiments, the final discovered equations included additional higher-order terms not shown in \cref{tab:cavity}. These terms typically have small coefficients and contribute marginally to the loss reduction. Therefore, they were truncated in the final summarization for clarity. It is also noteworthy that the coefficients of the shear-related terms identified by SITE exhibit slight deviations from those in the classical Newtonian constitutive model. These discrepancies, along with the emergence of higher-order terms, may be attributed to the fact that the simulated flows are not strictly within the continuum regime. Locally, some features of slip flow may appear. Based on the Knudsen number (Kn), flow regimes can be classified into continuum (Kn \textless 0.001), slip (0.001 \textless Kn \textless 0.1), transition (0.1 \textless Kn \textless 10), and free-molecular (Kn \textgreater 10) regimes \citep{reeseNewDirectionsFluid2003}. Since molecular simulations are computationally expensive in the fully continuum regime, we selected a global ${\text{Kn}} = 0.005$ to balance accuracy and efficiency. Consequently, local deviations from continuum assumptions may still exist.

\subsection{Discussion on magnitude disparities}
\begin{figure} 
\centering  
\includegraphics[width=13cm]{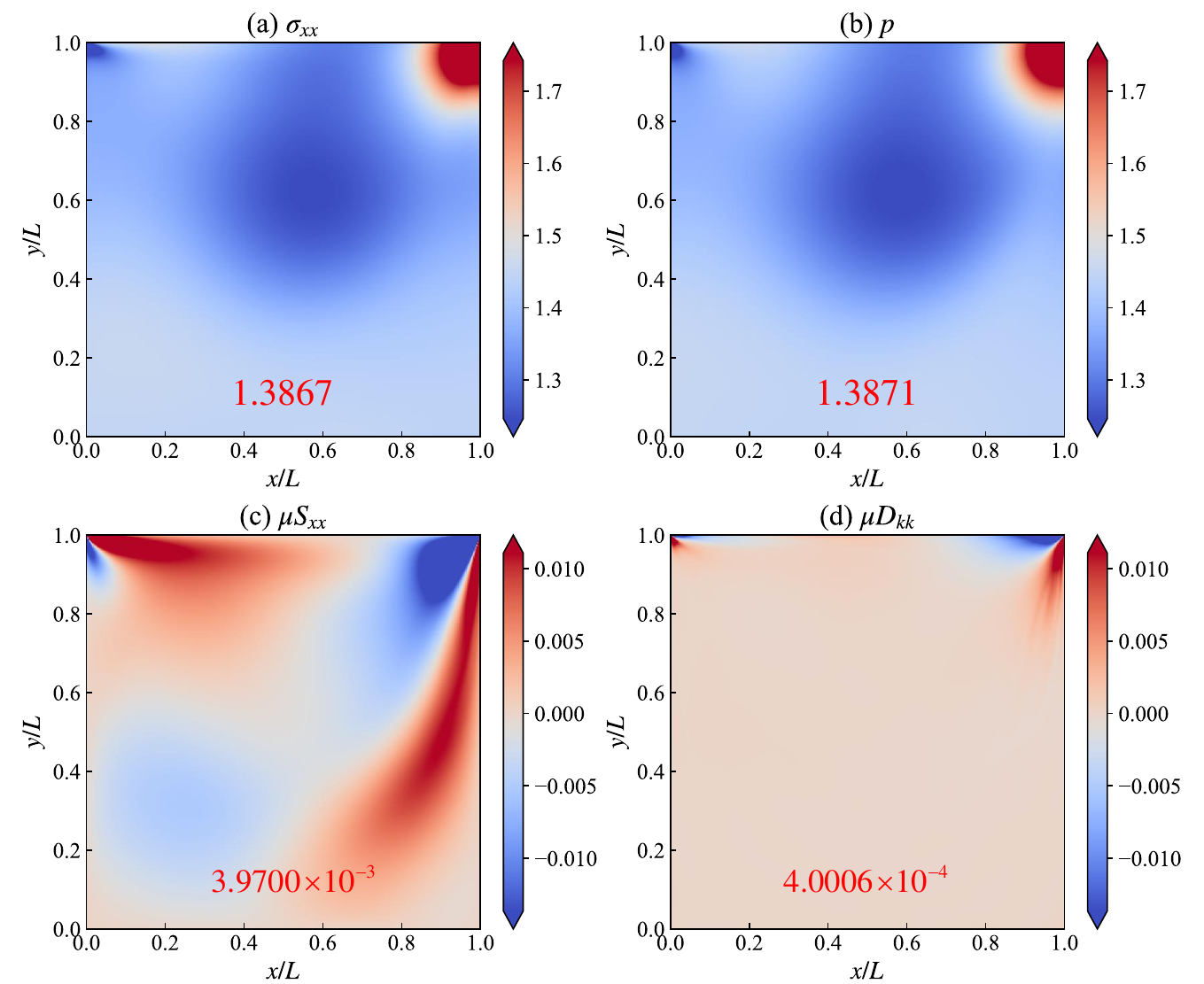}
\caption{Distributions of the constitutive terms in the xx-direction of the compressible cavity flow case. (a) Stress component ${\sigma _{xx}}$; (b) Static pressure $p$; These two quantities have similar magnitudes and are shown with a shared color scale. (c) Shear viscosity contribution $\mu {S_{xx}}$; (d) Compressible contribution $\mu {D_{kk}}$; These two quantities have similar magnitudes and are shown with a shared color scale. The average absolute values over the entire domain are indicated in red text within each subplot. Quantitatively, $\mu {D_{kk}}$ is roughly four orders of magnitude smaller than the dominant stress terms, posing a significant challenge for data-driven discovery.} 
\label{fig6}
\end{figure}

A key challenge in data-driven modeling lies in the simultaneous identification of terms with vastly different magnitudes from noisy data. In our compressible cavity flow case, this issue becomes particularly evident. The off-diagonal components, ${\sigma _{xy}}$ and ${\sigma _{yx}}$, is governed solely by shear deformation and is thus dominated by the symmetric part of the velocity gradient tensor. Consequently, identifying ${S _{ij}}$-related terms in this case is relatively straightforward. In contrast, the diagonal components of the stress tensor are composed of contributions from shear stress term, compressible term (linked to ${D _{kk}}$), and static pressure $p$. Among these, the compressible term typically has the smallest magnitude in the current case (see \cref{fig6}).

In this case, the problem can be simplified by subtracting the pressure to isolate the deviatoric stress ${\tau _{ij}} = {\sigma _{ij}} - p{\delta _{ij}}$. However, to rigorously evaluate whether SITE can discover small-magnitude terms amid much larger dominant contributions, we deliberately choose to identify the complete constitutive relation of $\sigma _{ij}$. The results in §\ref{sec:4.2} have shown that SITE is capable of reliably identifying the compressible term under such challenging conditions. This highlights the framework’s robustness to large disparities in variable magnitudes.

This robustness arises from SITE's design principles that explicitly address magnitude disparities without distorting physical significance. Compared to sparse regression-based methods which relies on standardized feature spaces and regularization to suppress terms with small numerical contributions, SITE operates directly in the native physical scale and preserves candidate terms through evolutionary mechanisms. Sub-expressions such as $\mu D_{kk}$ are retained and re-evaluated across generations, even if their immediate numerical contribution is minimal.

In addition to the overall magnitude disparity between different terms, another challenge specific to tensor equation identification lies in the magnitude disparity of different components within a single tensor. Unlike scalar regression tasks where each feature corresponds to a single physical quantity, tensor modeling requires treating entire tensor structures as regression targets. However, the magnitudes of different components within a tensor (e.g., the diagonal and off-diagonal terms in $\sigma_{ij}$) can vary significantly. This disparity can bias the learning process if not properly addressed. In SITE, this issue is mitigated by the design of optimization objective. The loss function (\ref{lossfunc}) incorporates a normalization scheme that rescales each component individually. This ensures that each tensor component contributes equally to the total loss, regardless of its absolute magnitude, thereby improving the robustness of coefficient identification.

\section{Conclusion}
\label{sec:conclusion}
This study introduces SITE, a symbolic regression framework tailored to discover governing equations involving second-order tensors. SITE addresses key limitations of conventional evolutionary approaches by enabling the identification of physically interpretable tensor equations with enhanced numerical efficiency.

To implement tensor symbolic regression, SITE leverages a host–plasmid structure and incorporates a genetic information retention strategy to enhance evolutionary robustness. A dimensional homogeneity check is enforced to ensure physical validity, while a seed injection strategy is adopted to overcome dimensional stagnation under strict constraints. Furthermore, a tensor linear regression technique replaces traditional random mutation–based coefficient optimization, enabling high efficiency and stability.

We validate SITE using two benchmark problems—including the Maxwell stress tensor and the Reynolds-stress transport equation—demonstrates its ability to recover ground-truth equations from noisy and limited data. Beyond synthetic tests, SITE is applied to DSMC-generated molecular data for lid-driven cavity flows, successfully identifying constitutive relations under both compressible and incompressible conditions. Notably, SITE also proves robustness in identifying small-magnitude terms, highlighting its sensitivity to meaningful physical features. 

In summary, SITE provides a general, flexible, and interpretable framework for tensor equation discovery in high-dimensional physical systems. While this work focuses on second-order tensors, future extensions may explore the incorporation of higher-order tensors and more generalized symbolic structures. Such advancements could significantly broaden the framework’s applicability both in science and engineering.

\backsection[Acknowledgements]{The authors thank P. Luan for his help with data generation using OpenFOAM.}

\backsection[Funding]{This work was supported by the National Natural Science Foundation of China (grant Nos. 12272028 and 92371102).}

\backsection[Declaration of interests]{The authors report no conflict of interest.}

\backsection[Data availability statement]{The datasets and source codes used in this work are available on GitHub at \href{https://github.com/PistilReaper/SITE}{https://github.com/PistilReaper/SITE}.}

\backsection[Author contributions]{T.C., W.M. and J.Z. contributed to the ideation and design of the research; T.C. and H.Y. generated the datasets; T.C. performed the research; T.C., H.Y. and J.Z. wrote the paper.}

\appendix
\renewcommand{\thetable}{A.\arabic{table}}
\setcounter{table}{0}

\section{Genetic operators}
\label{appA}
Within the evolutionary process of SITE, both the host population and the plasmid population are subjected to genetic operator applications. More specifically, the genetic operations on the host and plasmid populations are decoupled. First, the host population undergoes modifications and crossovers. After completing these operations on the host population, the plasmid population then undergoes modifications. Modification (\cref{alg:modification}) is divided into mutation and transposition, where transposition is further subdivided into gene fragment inversion, insertion sequence (IS) transposition, root insertion sequence (RIS) transposition, and whole gene transposition. Crossover (\cref{alg:crossover}) is divided into single-point crossover, double-point crossover, and whole gene crossover. These genetic operators are applied to the offspring with certain probabilities. 

\begin{algorithm}
\caption{Modification}
\label{alg:modification}
\begin{algorithmic}[1] 
\State \textbf{Input:} population, operator, $p_b$
\Function{apply\_modification}{population, operator, $p_b$}
\State $N \gets \text{length(population)}$ \Comment{Get the population size}
\For{$i \gets 0$ to $N-1$}
    \If{random() \textless $ p_b$}
        \State population[$i$] $\gets$ operator(population[$i$])
        \State delete fitness of population[$i$] \Comment{Invalidate fitness}
    \EndIf
\EndFor
\State \Return population
\EndFunction
\end{algorithmic}
\end{algorithm}

\begin{algorithm}
\caption{Crossover}
\label{alg:crossover}
\begin{algorithmic}[1]
\State \textbf{Input:} population, operator, $p_b$
\Function{apply\_crossover}{population, operator, $p_b$}
\State $N \gets \text{length(population)}$ \Comment{Get the population size}
\For{$i \gets 1$ to $N-1$ step $2$} \Comment{Iterate in pairs}
    \If{random() \textless $p_b$}
        \State $(\text{population}[i], \text{population}[i+1]) \gets$ \\
        \hspace{1.5em}operator(population[$i$], population[$i+1$])
        \State delete fitness of population[$i$]
        \State delete fitness of population[$i+1$]
    \EndIf
\EndFor
\State \Return population
\EndFunction
\end{algorithmic}
\end{algorithm}

In this work, the probabilities of applying each genetic operator refer to \citet{ferreiraGeneExpressionProgramming2001a}, listed in \cref{tab:genetic_operators}. It is important to note that, unlike other operators, the mutation operator's probability listed in the table is not its probability of being applied to an entire individual. In GEP, the mutation operator is applied to every offspring (i.e., the per-individual application probability is 1). Then, the operator traverses all gene elements in a chromosome, and each gene element is changed with the probability specified in the table (effectively, this is the calling probability for each gene element). As a result, even after the mutation operator is applied, an individual may remain unchanged if none of its gene elements are mutated. Additionally, genetic operators that operate directly on genes only become effective when the number of genes is greater than one.

\begin{table}
  \begin{center}
\def~{\hphantom{0}}
  \begin{tabular}{llc}
    Genetic operator & Object & Probability \\[6pt]
    Mutation                  & Host            & 0.2  \\
                              & Plasmid         & 0.05 \\[3pt]
    Inversion                 & Host            & 0.2  \\
                              & Plasmid         & 0.1  \\[3pt]
    IS transposition         & Host            & 0.2  \\
                              & Plasmid         & 0.1  \\[3pt]
    RIS transposition        & Host            & 0.2  \\
                              & Plasmid         & 0.1  \\[3pt]
    Gene transposition       & Host            & 0.2  \\
                              & Plasmid         & 0.1  \\[3pt]
    Single-point crossover   & Host            & 0.2  \\[3pt]
    Double-point crossover   & Host            & 0.2  \\[3pt]
    Gene crossover           & Host            & 0.2
  \end{tabular}
  \caption{Probabilities for genetic operators being invoked.}
  \label{tab:genetic_operators}
  \end{center}
\end{table}

In practical tests, SITE is not sensitive to the application probability of the genetic operators. Therefore, when choosing these probabilities, you can adjust them upward or downward according to whether you want to enhance exploration or achieve faster convergence.

\renewcommand{\thetable}{B.\arabic{table}}
\setcounter{table}{0}

\section{Hyperparameters}
\label{appB}
There are several key hyperparameters that govern SITE’s evolutionary process. \Cref{tab:site_hyperparameters} summarizes the values used in this study, which were selected based on prior benchmarking.

\begin{table}
  \begin{center}
\def~{\hphantom{0}}
  \begin{tabular}{lc}
    Hyperparameter & Value \\[6pt]
    Head length of the host gene       & 5    \\
    Head length of the plasmid gene    & 10   \\[3pt]
    Number of genes in a host CS       & 4    \\
    Number of genes in a plasmid CS    & 1    \\[3pt]
    Number of individuals in the population & 1600 \\
    The maximum number of generations  & 2000 \\[3pt]
    The number of elites               & 1    \\
    Tournament size                    & 200  \\
    The number of seed individuals     & 100
  \end{tabular}
  \caption{Values of hyperparameters used in SITE.}
  \label{tab:site_hyperparameters}
  \end{center}
\end{table}

The first four parameters are at the individual level. A larger head length allows for more diverse combinations of predefined candidate elements, which is particularly important when the terminal library is rich. The number of host genes directly controls the maximum number of independent terms allowed in the final constitutive relation derived via tensorial linear regression. For more complex problems, this value can be increased to capture additional physical mechanisms. The number of plasmid genes should not be excessive, as this would increase the computational burden associated with seed individual generation.

The remaining parameters are at the population evolution level. These settings generally do not have a significant impact on SITE’s optimization process. Compared to classical GEP settings, we have moderately increased the tournament size, which can enhance the convergence efficiency.

\section{Construction of the terminal library for constitutive relation identification}
\label{appC}
In constructing the candidate terminal library for the compressible cavity flow case, our objective was to avoid imposing any strong prior modeling assumptions regarding the form of the constitutive relation. Instead, we aimed to provide a physically comprehensive yet minimally biased set of inputs that could enable the discovery of the constitutive relation purely from data.

We initially included both the symmetric strain-rate tensor $S_{ij}$ and the antisymmetric rotation-rate tensor $\Omega_{ij}$ in the terminal library for tensors. However, preliminary experiments showed that $\Omega_{ij}$ was consistently absent in the expressions discovered by SITE. This result aligns with physical analysis: $\Omega_{ij}$ represents local rigid-body rotation, which does not contribute to momentum transport in isotropic fluids. As a result, the final terminal library for tensor is set to $\{ {S_{ij}},{\delta _{ij}}\} $.

For the scalar terminal library, we selected $\{ {D_{kk}},p,\mu \} $. Here, ${D_{kk}} = \nabla  \cdot u$ is the divergence of the velocity field and serves as a fundamental scalar invariant of the velocity gradient tensor ${D_{ij}}$. The static pressure $p$ and dynamic viscosity $\mu$ are intrinsic thermodynamic and transport properties of the flow, respectively. These scalar quantities enable SITE to capture both shear and compressible effects without presuming linear or Newtonian behavior.

\clearpage

\bibliographystyle{plainnat}
\bibliography{myreference}

\end{document}